\documentclass[aps,superscriptaddress,twocolumn,showpacs,pra]{revtex4-1}
\usepackage[utf8]{inputenc}
\usepackage{amsmath}
\usepackage{braket}
\usepackage{graphicx}
\usepackage{amsfonts}
\usepackage{pgfplots}
\usepackage{csquotes}
\usepackage{hhline}
\usepackage{amssymb}
\usepackage{listings}
\usepackage{color}

\makeatletter
\newcommand{\ssymbol}[1]{^{\@fnsymbol{#1}}}
\makeatother

\usepackage{lipsum}
\definecolor{codegreen}{rgb}{0,0.6,0}
\definecolor{codegray}{rgb}{0.5,0.5,0.5}
\definecolor{codepurple}{rgb}{0.58,0,0.82}
\definecolor{backcolour}{rgb}{0.95,0.95,0.92}
 
\lstdefinestyle{mystyle}{
    backgroundcolor=\color{backcolour},   
    commentstyle=\color{codegreen},
    keywordstyle=\color{magenta},
    numberstyle=\tiny\color{codegray},
    stringstyle=\color{codepurple},
    basicstyle=\footnotesize,
    breakatwhitespace=false,         
    breaklines=true,                 
    captionpos=b,                    
    keepspaces=true,                 
    numbers=left,                    
    numbersep=5pt,                  
    showspaces=false,                
    showstringspaces=false,
    showtabs=false,                  
    tabsize=2
}
 
\usepackage{subfigure}

\usepackage{dcolumn}
\usepackage{tabularx}
\usepackage[colorlinks=true,linkcolor=blue,citecolor=blue,urlcolor=blue]{hyperref}
\usepackage{longtable}
\usepackage{braket}
\usepackage{float}
\newcolumntype{C}{>{\centering\arraybackslash}X}

\begin{document}
\title{Direct generation of two-pair frequency entanglement via dual periodic poling in lithium niobate waveguides}

\author{Aakash Warke}
\email{warkeaakash@gmail.com}
\author{Krishna Thyagarajan}
\email{krishna.thyagarajan@bennett.edu.in}
\affiliation{Department of Physics, Bennett University, Greater Noida 201310, India}

\begin{abstract}
In this paper, we address the generation of a two-pair frequency entangled state using type-0 spontaneous parametric down-conversion process in a dual periodically poled lithium niobate waveguide. We show that, by suitable domain engineering with two periods of quasi-phase matching grating, it is possible to achieve a frequency entangled state with two different pairs of frequencies. Numerical simulations show that the output state can be maximally entangled. We also perform numerical simulations to address other interesting and useful entangled states which can be generated with the help of our scheme. The proposed scheme can help create efficient photonic setups required for quantum communication systems and can have various applications considering the increasing interests of energy-time entanglement in quantum information regime.
\end{abstract}

\begin{keywords}{Quantum optics, Waveguide optics, Quantum communication, Integrated optics}\end{keywords}

\maketitle
\section{Introduction}
The foundation of quantum mechanics relies on the possibility of superposition and entanglement \cite{AEPRA1935, JYPRL2013}. These phenomena form the basis for many applications originating from quantum mechanics such as quantum computation, quantum communication and more \cite{PBJSP1980,RFIJTP1982,KLNat2001,BBPRL1993,PPRA2000,CBSp1992}. Hence, the generation of entangled states is of crucial importance to such applications and to understand the foundations of quantum mechanics. The quantum communication schemes encompass photon entanglement to perform distinct communication tasks. This is the reason why entangled photon pairs have attracted so much attention over the past years \cite{YLPRA2016,MKNat2020,PKPRL1995,PKPRA1999,YCNat2016,FSOE2019,DONPJ2020, YWAPR2021}. The process of spontaneous parametric down-conversion employs the use of $\chi^{(2)}$ nonlinearity in crystals such as lithium niobate to down-convert incident light of certain frequency, namely pump, into two lower frequencies, signal and idler. This is possible through certain phase matching conditions which are achieved via periodic poling in these crystals. Multiple periods allow the simultaneous existence of the down-converted pairs of photons, which in turn help us generate entangled photon pairs \cite{STOSA2002,SLOSA2013}. This is a probabilistic way of generating entangled photons. They can exist in different degrees of freedom such as polarization, spatial modes, frequency etc. and a lot of research goes in exploring such quantum states, for it not only has applications but also helps us in understanding quantum mechanics at a deeper level \cite{KTPRA2009,VKSPRA2020,JLOL2012,TJEPRL2015,DBPRA2015,LOPRA2010,CBPRA2013,HKOL2015,FMOE2016}. Generation of hybrid- or hyper-entangled photon pairs has also attracted a lot of attention over the years due to its several applications in quantum technologies \cite{HJNat2014,DBJOSA2018,FGPRR2020,JLJOSA2013}. The use of waveguides for this purpose of generating entanglement gives an edge as it leads to augmented nonlinear efficiencies due to longer interaction lengths and tight confinement of interacting waves. The periodically poled lithium niobate is an essential substrate as it offers the largest nonlinear coefficient $d_{33}$, thus leading to highest nonlinearity and ultimately, effective entangled photon pairs at the output \cite{KTPRA2009}. Due to such high nonlinear coefficients, various schemes for generation of entangled photons have been reported by taking periodically poled lithium niobate (PPLN) or periodically poled potassium titanyl phosphate (PPKTP) waveguides into consideration \cite{HJPRL2014,OAJO2016,JZPRL2020,ZHLOE2011}.

Quantum frequency combs form a particularly useful resource for parallel quantum communication processes. These require discrete modes of frequencies. One way to deal with frequency's continuous behavior is to discretize it and generate biphoton combs entangled in frequency modes \cite{MKNat2019}. These kinds of states were first explored using the spontaneous parametric down-conversion (SPDC) process in bulk systems. They have also been examined for the higher dimensional case using SPDC \cite{YJPRL2003,ZXNat2015,RJQST2016}. Frequency combs were also generated using integrated optical micro-resonators owing to spontaneous four-wave mixing \cite{FMOE2016}. Frequency combs are characterized by closely lying frequencies. High dimensional frequency entangled states have been generated and exploited for coherent manipulation in Hydex and silicon nitride micro-ring resonators \cite{MKNat2017,PIOE2018}. More recently, a method was proposed to control the comb's symmetry by combining the cavity's spectral filtering effect with the control of temporal delay between the photons corresponding to a pair \cite{GMNPJ2020}. The reported results are undoubtedly of prime interest to us, but these solutions are based on setups that require demanding stabilization and strict control. 

In this paper, we present a scheme to directly generate two-pair frequency entanglement which is the fundamental criteria for generation of frequency combs, in a titanium in-diffused periodically poled lithium niobate waveguide via type-0 spontaneous parametric down-conversion process. We make this possible by suitable engineering of the quasi-phase matching (QPM) grating for two periods which satisfy both the SPDC processes simultaneously. One period corresponds to down-conversion of the pump light to signal and idler photons of certain frequencies whereas the other period corresponds to down-converting the same pump into a different set of frequencies. The output state will be thus entangled in two pairs of frequencies. It is then possible to have a beam splitter which transmits the signal wavelengths and reflects the idler wavelengths. Furthermore, it is also possible to separate all the four wavelengths into four ports leading to path entangled outputs. Compared to the case of frequency combs, here the entangled frequencies can be quite different. This can be interesting in cases where one of the entangled photons is used for atom-photon interaction while the other photon is sent over an optical fiber channel.

We assume extraordinary polarization for the pump light, signal and idler photons. This will allow us to access the $d_{33}$ element of lithium niobate leading to an effective output entangled state as discussed earlier. We discuss interesting implications of considering different polarizations for this process later on in the paper. This scheme should be of great interest to applications requiring frequency entangled photon pairs, for instance, frequency-multiplexed optical quantum information processing, and can even be realized by commercially available optical components.

The paper is arranged as follows. In Section \ref{MP1_Sec2}, we give a quantum mechanical analysis of the two SPDC processes to describe the generation of a two-pair frequency entangled state. In Section \ref{MP1_Sec3} we present the numerical results of simulations carried out by modeling  the structure of our considered titanium in-diffused channel waveguide, addressing the possibility to generate a maximally entangled state. This is done with the intention of providing the practicality of this idea. We then discuss the issues pertaining to bandwidth of the two coupled interaction processes. We also discuss different entangled states that can be generated through our proposed scheme and perform numerical simulations to provide their practicality. In Section \ref{MP1_Sec4}, we give a brief conclusion of this paper.

\section{Quantum Mechanical Analysis of SPDC for the two interaction processes \label{MP1_Sec2}}
We consider the process of parametric down-conversion in a titanium in-diffused channel waveguide consisting of a z-cut, x-propagating lithium niobate as its substrate. This substrate is poled as per the functional dependence of nonlinear coefficient $d_{33}$ to satisfy both the required QPM conditions simultaneously. We assume all the interacting modes involved in the process to be extraordinarily polarized. It is therefore a type-0 spontaneous parametric down-conversion process. As light of wavelength $\lambda_p$ is launched into the crystal, it down-converts to photon pairs of different wavelengths but the QPM conditions decide which of them should be more efficient at the output. The objective is to generate an entangled state given by:
\begin{equation}
    \Ket{\Psi}=\int d\omega_{s1}C_1\Ket{\omega_{s1},\omega_{i1}}+\int d\omega_{s2}C_2\Ket{\omega_{s2},\omega_{i2}}
    \label{MP1_Eq1}
\end{equation}
Here, $\omega_{s1}$, $\omega_{s2}$ represent the signal frequencies and $\omega_{i1}$, $\omega_{i2}$ represent the idler frequencies of the two down-converted pairs. $C_1$ and $C_2$ are the amplitudes of the respective down-conversion processes. In the case of a maximally entangled state, these two values need to be equal. One period generates the pair $\Ket{\omega_{s1},\omega_{i1}}$, and the other period generates $\Ket{\omega_{s2},\omega_{i2}}$ from the incident pump light. It is necessary that these conversions follow energy conservation which is given by:
\begin{eqnarray}
    \omega_p=\omega_{s1}+\omega_{i1}\nonumber\\
    \omega_p=\omega_{s2}+\omega_{i2}
    \label{MP1_Eq2}
\end{eqnarray}
First, we consider the electric field profiles of each of the four modes, derive the interaction Hamiltonian using the $\chi^{(2)}$ nonlinearity, and then use this Hamiltonian to derive the output state \cite{KTPRA2009}.
Power of the incident pump light in this process is usually high and thus, its electric field can be assumed to be represented by a classical field. The signal and idler fields are quantized and are therefore represented by their respective operators. The electric field distributions for the pump and four interacting modes are thus given by:
\begin{equation}
    \vec{E}_{\omega_p}=\frac{1}{2}e_{p}(\vec{r})E_p\big[e{^{i(k_px-\omega_pt)}}+e{^{-i(k_px-\omega_pt)}}\big]\hat{z}
    \label{MP1_Eq3}
\end{equation}

\begin{eqnarray}
    \hat{E}_{\omega_{m}}=i\int{d\omega_me_{\omega_m}(\vec{r})\sqrt{\frac{\hbar\omega_m}{2\epsilon_mL}}}\big(\hat{a}_{\omega_m}e^{i(k(\omega_m)x-\omega_mt)}\nonumber\\\nonumber\\
    -\hat{a^{\dagger}}_{\omega_m}e^{-i(k(\omega_m)x-\omega_mt)}\big)\hat{z}\hspace{17pt}
    \label{MP1_Eq4}
\end{eqnarray}
The subscript $p$ stands for denoting pump in Eq.\eqref{MP1_Eq3}. In Eq. \eqref{MP1_Eq4}, $m \in \{s1,i1,s2,i2\}$ and $\omega_m$ corresponds to the four frequencies taken into consideration for the four discrete modes. $e_{p}(\vec{r}),e_{\omega_m}(\vec{r})$ represent the transverse dependence of these
modal fields corresponding to extraordinary polarization of the considered pump and the
four frequencies. The length of interaction is represented by $L$ and $\epsilon_m$ represents the optical dielectric permittivity of the medium. Annihilation and creation operators of
frequencies into their respective modes are represented by $\hat{a}_{\omega_m}$ and $\hat{a^{\dagger}}_{\omega_m}$. \\
Quasi-phase matching (QPM) ensures that the entanglement between the four modes which will encompass these four discrete frequencies is generated efficiently. These QPM periods are given by $K_1$ and $K_2$:
\begin{eqnarray}
    K_1 = \frac{2\pi}{\Lambda_1}=2\pi\Big\{\frac{n_p}{\lambda_p}-\frac{n_{s1}}{\lambda_{s1}}-\frac{n_{i1}}{\lambda_{i1}}\Big\}
    \label{MP1_Eq5}
\end{eqnarray}
\begin{eqnarray}
    K_2 = \frac{2\pi}{\Lambda_2}=2\pi\Big\{\frac{n_p}{\lambda_p}-\frac{n_{s2}}{\lambda_{s2}}-\frac{n_{i2}}{\lambda_{i2}}\Big\}
    \label{MP1_Eq6}
\end{eqnarray}
Here, $n_p,n_{s1},n_{i1},n_{s2},n_{i2}$ represent the effective indices of the modes with extraordinary polarization at wavelengths $\lambda_p,\lambda_{s1},\lambda_{i1},\lambda_{s2},\lambda_{i2}$ respectively and are calculated using the Sellmeier’s equations \cite{DEJOSA1997}. Now, using the nonlinear polarization tensor, the total electric field components along x, y and z axes are given by:
\begin{eqnarray}
    E_1=0\nonumber\hspace{60pt}\\
    E_2=0\nonumber\hspace{60pt}\\
    E_3=E_{\omega_p}+E_{\omega_{s1}}+E_{\omega_{i1}}+E_{\omega_{s2}}+E_{\omega_{i2}}
    \label{MP1_Eq7}
\end{eqnarray}

\begin{figure}[]
    \includegraphics[scale=0.3]{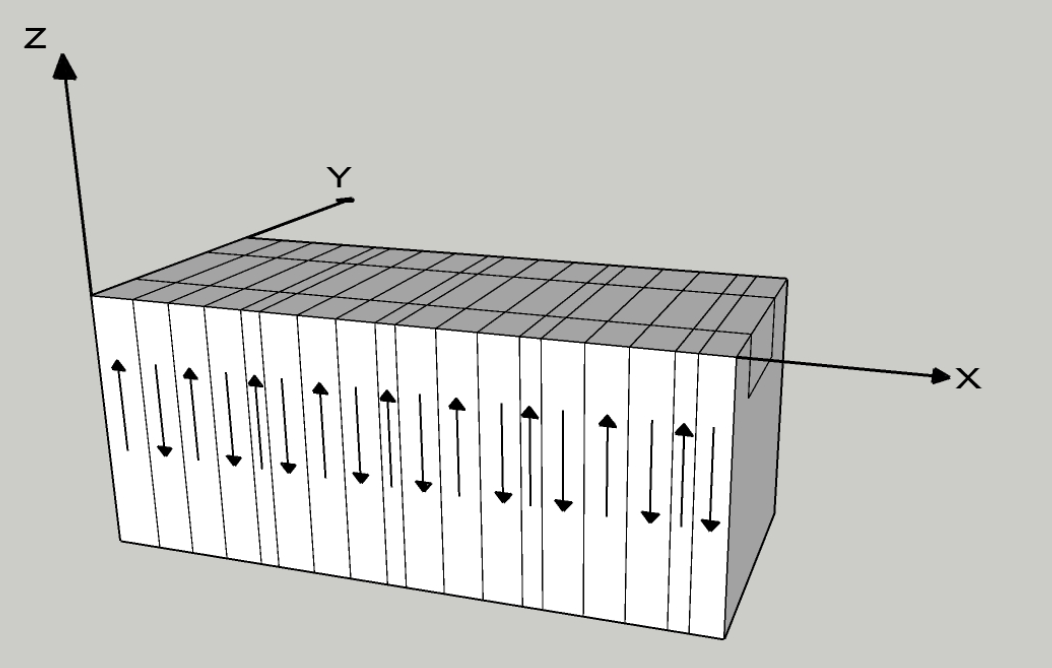}
    \caption{This figure depicts the waveguide geometry for generating the two-pair frequency entangled state as shown in Eq. \eqref{MP1_Eq1}, using two SPDC interaction processes that satisfy the QPM conditions mentioned in Eq. \eqref{MP1_Eq5} and \eqref{MP1_Eq6}. The optic axis of this lithium niobate substrate is along the z direction. The electric field profiles $\vec{E}_{\omega_p}$, $\hat{E}_{\omega_{s1}}$, $\hat{E}_{\omega_{i1}}$, $\hat{E}_{\omega_{s2}}$ and $\hat{E}_{\omega_{i2}}$ are all along the optic axis as they are extraordinarily polarized.}
    \label{MP1_Fig1}
\end{figure}

The subscript indices 1,2, and 3 refer to components along x, y, and z axes as shown in Fig. \ref{MP1_Fig1}. Since this happens to be a type-0 down-conversion process in a lithium niobate crystal, the $d_{33}$ coefficient is involved for derivation of the Hamiltonian \cite{JLPRA2011}. Taking the expressions for electric field profiles given in Eq. \eqref{MP1_Eq3} and \eqref{MP1_Eq4}, the interaction Hamiltonian is derived as shown in Eq. \eqref{MP1_Eq8} \cite{KTPRA2009}. 

\begin{equation}
    \hat{H}_{int} = -4\epsilon_0\int{ d_{33}\big(\vec{E}_{\omega_p}\hat{E}_{\omega_{s1}}\hat{E}_{\omega_{i1}}+\vec{E}_{\omega_p}\hat{E}_{\omega_{s2}}\hat{E}_{\omega_{i2}}\big)d^3r}
    \label{MP1_Eq8}
\end{equation}

We assume a dual periodic poling \cite{KTPRA2009} and replace $d_{33}$ in Eq. \eqref{MP1_Eq8}'s interaction Hamiltonian by:
\begin{equation}
    \bar{d}_{33} = \frac{-4d_{33}}{\pi^2}\big(e^{iK_1x}+e^{-iK_1x}-e^{iK_2x}-e^{-iK_2x}\big)
\end{equation}
Now, substituting the expressions for respective electric field profiles, and applying rotating-wave approximation, we derive the interaction Hamiltonian which is given as:

\begin{eqnarray}
    \hat{H}_{int} = \int\int d\omega_{s1} d\omega_{i1} C_1^{(1)}\big(\hat{a^{\dagger}}_{\omega_{s1}}\hat{a^{\dagger}}_{\omega_{i1}}e^{-i(\omega_p-\omega_{s1}-\omega_{i1})t}\nonumber\\
    + h.c.\big) \hspace{20pt}\nonumber\\ +
    \int\int d\omega_{s2} d\omega_{i2} C_2^{(1)}
    \big(\hat{a^{\dagger}}_{\omega_{s2}}\hat{a^{\dagger}}_{\omega_{i2}}e^{-i(\omega_p-\omega_{s2}-\omega_{i2})t}\nonumber\\
    + h.c. \big)\hspace{20pt}
    \label{MP1_Eq10}
\end{eqnarray}
\begin{eqnarray}
    C_1^{(1)} = -\Bigg(\frac{4d_{33}E_p\hbar I_1\sqrt{\omega_{s1}\omega_{i1}}}{\pi^2 n_{s1}n_{i1}}\Bigg)\exp{\Big(-\frac{i\Delta k_1 L}{2}}\Big)\nonumber\\sinc\Big(\frac{\Delta k_1 L}{2}\Big)\hspace{12pt}
    \label{MP1_Eq11}
\end{eqnarray}
\begin{eqnarray}
    C_1^{(2)} = -\Bigg(\frac{4d_{33}E_p\hbar I_2\sqrt{\omega_{s2}\omega_{i2}}}{\pi^2 n_{s2}n_{i2}}\Bigg)\exp{\Big(-\frac{i\Delta k_2 L}{2}}\Big)\nonumber\\
    sinc\Big(\frac{\Delta k_2 L}{2}\Big)\hspace{12pt}
    \label{MP1_Eq12}
\end{eqnarray}
Here, h.c. refers to the Hermitian conjugate and $\Delta k_{1(2)}$ represents the phase mismatch given by:\\
\begin{eqnarray}
    \Delta k_1 = K_1 - 2\pi\big\{\frac{n_p}{\lambda_p}-\frac{n_{s1}}{\lambda_{s1}}-\frac{n_{i1}}{\lambda_{i1}}\big\}\\
    \Delta k_2 = K_2 - 2\pi\big\{\frac{n_p}{\lambda_p}-\frac{n_{s2}}{\lambda_{s2}}-\frac{n_{i2}}{\lambda_{i2}}\big\}
\end{eqnarray}

In Eq. \eqref{MP1_Eq11} and \eqref{MP1_Eq12}, $I_1$ and $I_2$ denote the overlap integrals which are crucial in determining the overlap between the interacting modes, which decide whether the state can be maximally entangled or not. We discuss more about this in Section \ref{MP1_Sec3}. These overlap integrals are given by:
\begin{eqnarray}
    I_1 = \int\int e_{p}(\vec{r})e_{s1}(\vec{r})e_{i1}(\vec{r})dydz\nonumber\\
    I_2 = \int\int e_{p}(\vec{r})e_{s2}(\vec{r})e_{i2}(\vec{r})dydz
    \label{MP1_Eq15}
\end{eqnarray}
Using the Hamiltonian shown in Eq. \eqref{MP1_Eq11}, we calculate the output state as:
\begin{equation}
    \Ket{\Psi}_{out} = e^{-i\hat{H}_{int}t/\hbar}\Ket{0,0}
    \label{MP1_Eq16}
\end{equation}
Taking Eq. \eqref{MP1_Eq16}, using the energy conservation, and substituting the values of respective parameters, we get the following output state:
\begin{eqnarray}
    \Ket{\Psi}=\int d\omega_{s1}C_1\Ket{\omega_{s1},\omega_{i1}}\nonumber\\
    +\int d\omega_{s2}C_2\Ket{\omega_{s2},\omega_{i2}}
    \label{MP1_Eq17}
\end{eqnarray}
where,
\begin{eqnarray}
    C_1 = -\frac{it}{\hbar}C_1^{(1)}\nonumber\\
    C_2 = -\frac{it}{\hbar}C_1^{(2)}
\end{eqnarray}\\
As one can observe, the state $\Ket{\Psi}$ is non-separable and thus, entangled. Therefore, we have achieved generating entanglement between two pairs of different frequencies using the aforementioned geometry in Fig. \ref{MP1_Fig1} of a titanium-indiffused channel waveguide with lithium niobate as its substrate. Depending on the overlap integrals and effective indices of the interacting modes of the four frequencies, we can conclude whether the state can be maximally entangled. As mentioned earlier, it is possible to now use a beam splitter for the purpose of separation by transmitting frequencies of the first mode of each pair, and reflecting the frequencies corresponding to the second mode. It is also possible to separate all the four frequencies into four ports which leads to a path entangled output. In Section \ref{MP1_Sec3}, we consider practical values of various parameters involved in the process and show that by properly designing the geometry of the waveguide, it is indeed possible to achieve a maximally entangled state.

\section{Numerical Simulations and Discussion}\label{MP1_Sec3}
For numerical simulation, we consider a titanium in-diffused channel waveguide and follow the procedure given in Ref. \cite{KTPRA2009}. We find the parameters $\alpha_{ym}$ and $\alpha_{zm}$, $m \in \{s1,i1,s2,i2\}$, by maximizing the expression for $n_{eff}$. Using these parameters, we derive the analytical expressions for $e_p(\vec{r})$, $e_{\omega_{s1}}(\vec{r})$, $e_{\omega_{i1}}(\vec{r})$, $e_{\omega_{s2}}(\vec{r})$ and $e_{\omega_{i2}}(\vec{r})$. We then solve the two overlap integrals given in \eqref{MP1_Eq15}. The value of $n_{eff}$ gives the effective extraordinary indices of the propagating pump and four frequency modes along the waveguide. These indices are then used to obtain the QPM periods. Combining all of this together, we can calculate the factor which relates $C_1^{(1)}$ and $C_1^{(2)}$ given in Eq. \eqref{MP1_Eq11} and \eqref{MP1_Eq12} by the following expression \cite{KTPRA2009}:
\begin{equation}
    \gamma = \frac{min(C_1^{(1)},C_1^{(2)})}{max(C_1^{(1)},C_1^{(2)})}
\end{equation}\\

$\gamma$ is referred to as the degree of entanglement. As it is observed, $n_{eff}$ is a function of width and height/depth of the waveguide, which means that the QPM period will also be a function of these parameters of the waveguide. This implies that the degree of entanglement will depend on the width and depth of the waveguide. Therefore, by considering different values of width and depth, we can determine which set of periods is the most suitable to make sure $\gamma$ approaches to 1, leading to a maximally entangled state. The values of wavelengths we choose for the pump and four discrete modes are represented in Table \ref{MP1_Table1} along with their respective extraordinary waveguide index variations.

\begin{table}
\centering
\begin{tabular}{|c c c c c c c c c|} 
\hline & & & & & & & &\\
& $Mode$ & & & $\lambda (nm)$ & & & $\Delta n_e$ &\\
& & & & & & & &\\
\hline 
& & & & & & & &\\
& $\lambda_p$ & & & 519 & & &  0.0037 &\\
& $\lambda_{s1}$ & &  & 780 & & & 0.0030 &\\
& $\lambda_{i1}$ & & & 1551.03 & & & 0.0025 &\\
& $\lambda_{s2}$ & & & 775 & & & 0.0030 &\\
& $\lambda_{i2}$ & & & 1571.19 & & & 0.0025 &\\
& & & & & & & &\\
\hline
\end{tabular}
\caption{This table represents the values of considered wavelengths corresponding to the pump and the four modes of the entangled state, and their extraordinary waveguide index variation because of titanium in-diffusion.}
\label{MP1_Table1}
\end{table}
\begin{table}
\centering
\begin{tabular}{|c c c c c c c c c c c c c c c|} 
\hline & & & & & & & & & & & & & &\\
& $Depth (\mu m)$ & & & $Width (\mu m)$ & & & $\gamma$ & & & $\Lambda_1 (\mu m)$ & & & $\Lambda_2 (\mu m)$ & \\
& & & & & & & & & & & & & &\\
\hline & & & & & & & & & & & & & &\\
& $6.5$ & & & $6.5$ & & & 0.9351 & & & 6.7874 & & & 6.8228 &\\
& $8$ & & & $8$ & & & 0.9750 & & & 6.7903 & & & 6.8251 &\\
& $10$ & & & $10$ & & & 0.9817 & & & 6.7969 & & & 6.8316 &\\
& $12$ & &  & $12$ & & & 0.9842 & & & 6.8027 & & & 6.8374 &\\
& & & & & & & & & & & & & &\\
\hline
\end{tabular}
\caption{This table represents the variation of the degree of entanglement and the two QPM periods for different depths and widths of the titanium in-diffused waveguide with lithium niobate as its substrate for the entangled state represented by Eq. \eqref{MP1_Eq17}. $\gamma$ approaches 1 for a maximally entangled quantum state.}
\label{MP1_Table2}
\end{table}

\begin{figure}[]
    \includegraphics[scale=0.285]{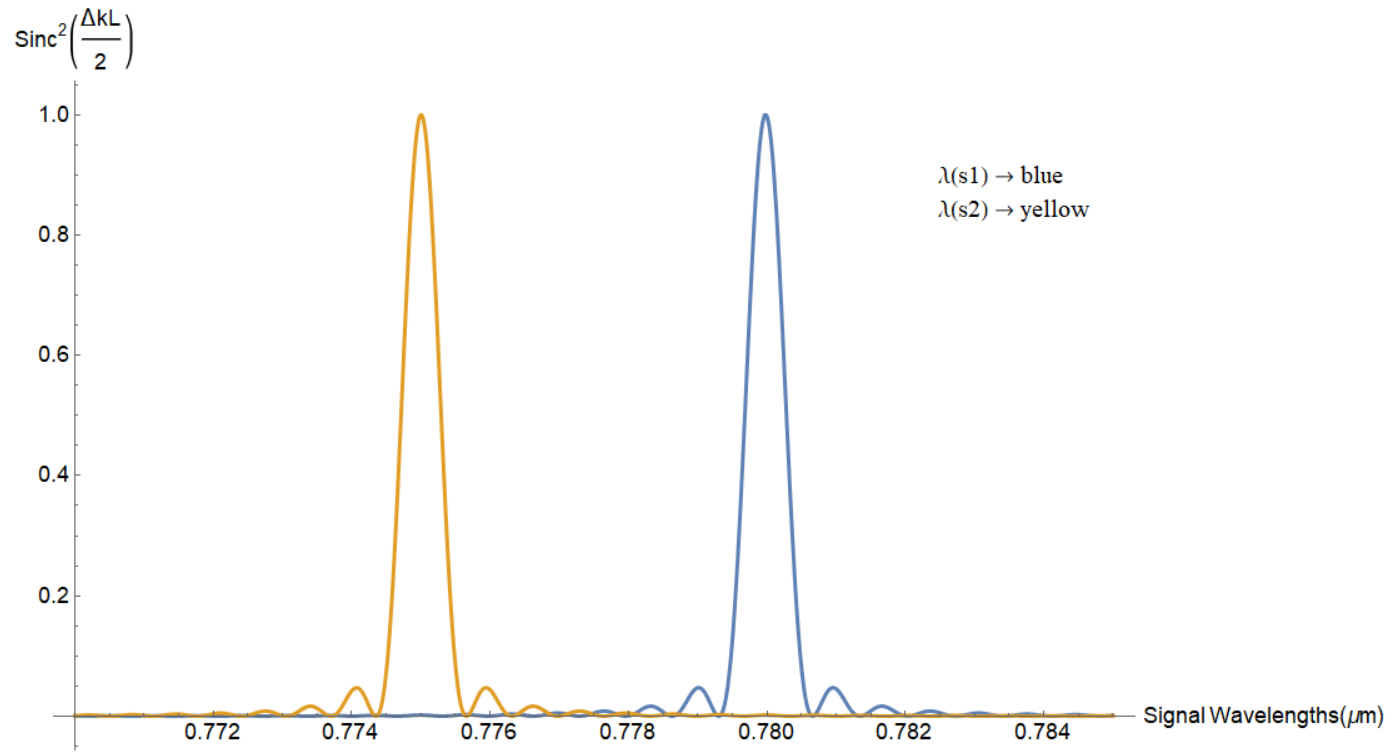}
    \caption{This figure represents the spectra corresponding to type-0 SPDC processes to generate the entangled state represented in Eq. \eqref{MP1_Eq17} for considered signal wavelengths of $\lambda_{s1}$ and $\lambda_{s2}$, which are 780 nm and 775 nm respectively. For a centimeter long interaction ($L$ = 1 cm) with a waveguide of $w = h = 10 \mu m$, the bandwidth of the process corresponding to $\lambda_{s1}$ is found to be 1.467 nm and the bandwidth of the process corresponding to $\lambda_{s2}$ is found to be 1.407 nm.}
    \label{MP1_Fig2}
\end{figure}

\begin{figure}[]
    \includegraphics[scale=0.26]{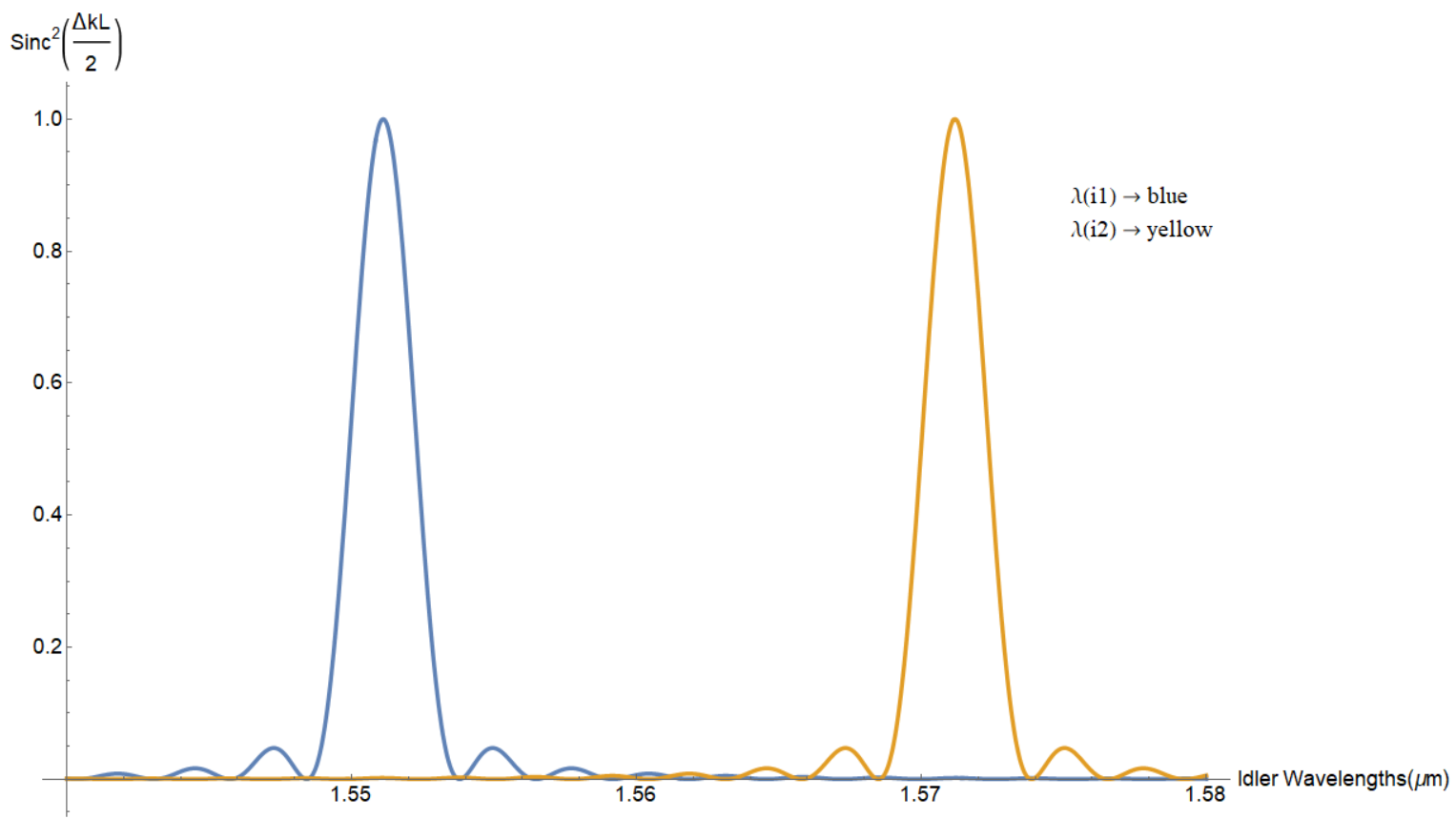}
    \caption{This figure represents the calculated spectra for the SPDC processes corresponding to idler wavelengths, $\lambda_{i1}$, $\lambda_{i2}$, which are 1.55103 $\mu m$ and 1.5719 $\mu m$ respectively. The SPDC process corresponds to the generation of entangled state represented in Eq. \eqref{MP1_Eq17} with $w$ = $h$ = 10 $\mu$m and $L$ = 1 cm. Bandwidths of $\lambda_{i1}$ and $\lambda_{i2}$ are found to be 5.799 nm and 5.788 nm respectively.}
    \label{MP1_Fig3}
\end{figure}

Table \ref{MP1_Table2} refers to the data acquired after performing the numerical simulations by taking into account the state given in Eq. \eqref{MP1_Eq17}. As inferred, it is possible to vary the parameters of the waveguide such that the state is maximally entangled. Fig \ref{MP1_Fig2} represents the spectrum corresponding to signal wavelengths $\lambda_{s1}$ and $\lambda_{s2}$. It is observed that outputs are non-overlapping and the bandwidth calculated for $\lambda_{s1}$ for its respective down-conversion process is 1.467 nm whereas the bandwidth calculated for $\lambda_{s2}$ with its corresponding spectra is 1.407 nm. Fig. \ref{MP1_Fig3} represents the spectra of the interacting modes corresponding to idler wavelengths.\\

It is interesting to note that if one wants to choose between more than two, say three pairs of frequencies, then three quasi-phase matching periods will have to be satisfied with appropriate consideration of wavelengths and waveguide parameters. However, the more pairs we look forward to choose from, lower the efficiency would be. Hence, different waveguide geometries can be used to obtain a favorable output, for instance, in this case, a three-waveguide directional coupler could be feasible \cite{DBPRA2015}.\\

As mentioned before, we have considered extraordinary polarization for all the frequencies involved in the process. If we are to consider the
polarization of the propagating waves in the medium for the geometry mentioned in Fig. \ref{MP1_Fig1}, interesting entangled states can be generated. For example, consider the state in Eq. \eqref{MP1_Eq20}.

\begin{eqnarray}
    \Ket{\Psi}_1=\int d\omega_{s1o}C_1\Ket{\omega_{s1o},\omega_{i1e}}\nonumber\\
    +\int d\omega_{s2e} C_2\Ket{\omega_{s2e},\omega_{i2o}}
    \label{MP1_Eq20}
\end{eqnarray}
Here, the subscripts $o$, $e$ refer to ordinary and extraordinary polarizations of the four interacting frequency modes respectively. Cross-polarization implies that it will take a type-II down-conversion process to generate this entangled state. Such an entangled state will help acquire different frequencies, only by measuring once, and thus can have numerous applications. In order to provide the practicality of such a state, we formulate the QPM periods analogously, and then perform numerical simulation following Ref. \cite{KTPRA2009}. We assume the pump to be of ordinary polarization and evaluate the effective indices for wavelengths shown in Table \ref{MP1_Table1}. Corresponding waveguide index variations for ordinary polarization are shown in Ref.\cite{KTPRA2009}. Table \ref{MP1_Table3} shows the results of the variation of degree of entanglement and poling periods for different widths and depths of the waveguide for the entangled state given in \eqref{MP1_Eq20}. 

\begin{figure}[]
    \includegraphics[scale=0.3]{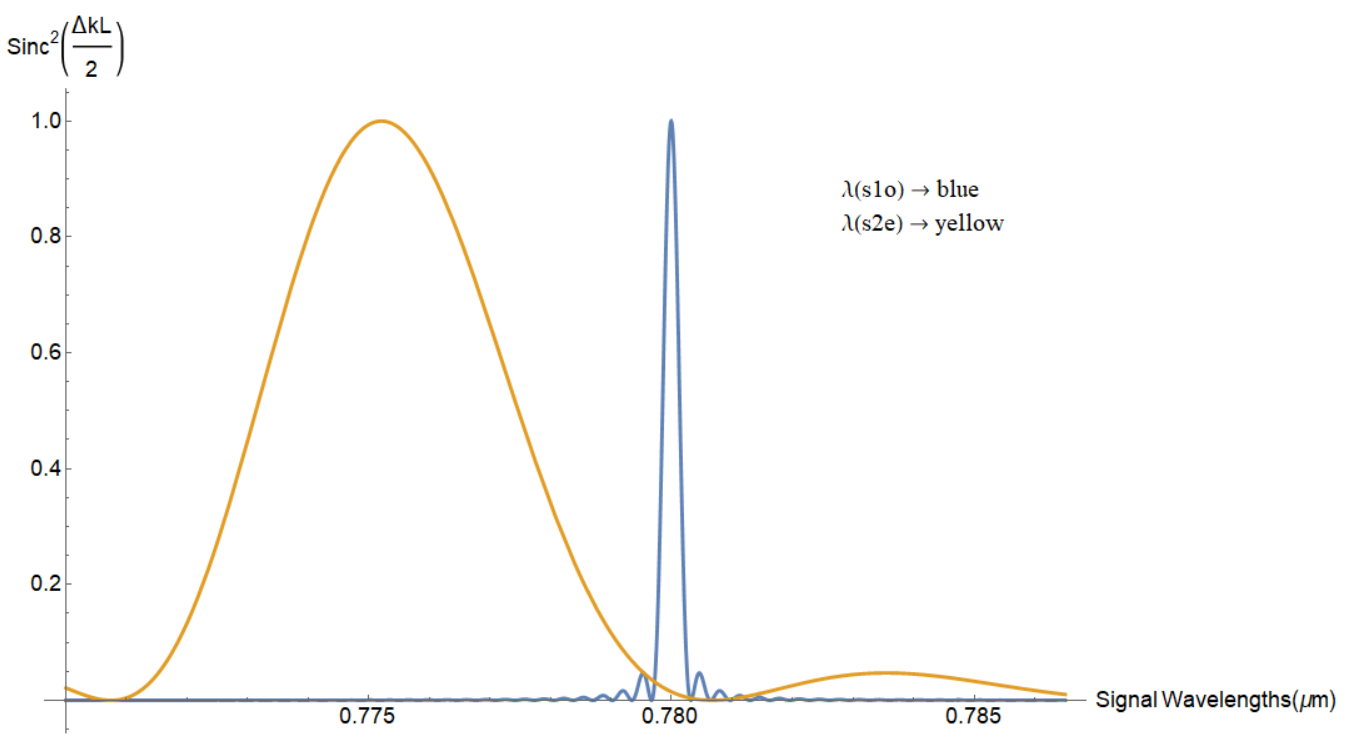}
    \caption{This figure represents the calculated spectra of type-II SPDC processes for the generation of entangled state represented in Eq. \eqref{MP1_Eq20}. The considered wavelengths, $\lambda_{s1o}$ and $\lambda_{s2e}$, are 780 nm and 775 nm respectively. For a centimeter long interaction ($L$ = 1 cm) with $w = h = 6.5 \mu m$, the bandwidth of the process corresponding to $\lambda_{s1o}$ is 0.499 nm and the bandwidth of the process corresponding to $\lambda_{s2e}$ is 1.952 nm. There seems to be a slight overlap in the two plots, but it is not a significant overlap as $\lambda_{s1o} - \Delta\lambda_{s1o}$ lies outside the range of $\lambda_{s2e} + \Delta\lambda_{s2e}$ and hence, can be differentiated experimentally \cite{KTPRA2009}. $\Delta\lambda_{s1o}$ and $\Delta\lambda_{s2e}$ represent the bandwidths of the two interaction processes.}
    \label{MP1_Fig4}
\end{figure}

\begin{figure}[]
    \includegraphics[scale=0.3]{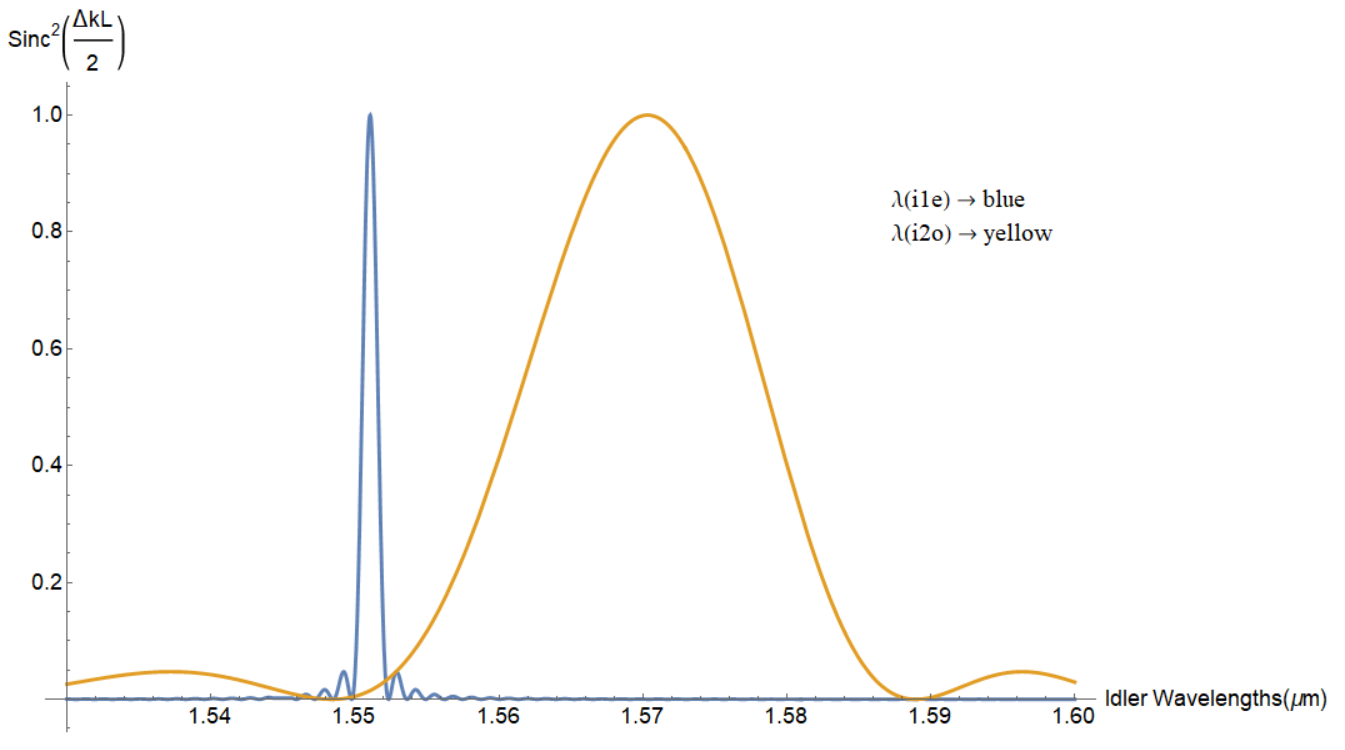}
    \caption{This figure represents the bandwidth calculations of the SPDC processes corresponding to idler wavelengths, $\lambda_{i1e}$ and $\lambda_{i2o}$ of 1.55103 $\mu m$ and 1.5719 $\mu m$ respectively to generate the entangled state represented in Eq. \eqref{MP1_Eq20}. For a centimeter long interaction ($L$ = 1 cm) with $w = h = 6.5 \mu m$, bandwidth corresponding to $\lambda_{i1e}$ is found to be 1.972 nm and that of $\lambda_{i2o}$ is found to be 8.032 nm. The slight overlap in this case is not significant due to the same reason mentioned in the explanation of Fig. \ref{MP1_Fig4}.}
    \label{MP1_Fig5}
\end{figure}

\begin{table}
\centering
\begin{tabular}{|c c c c c c c c c c c c c c c|} 
\hline & & & & & & & & & & & & & &\\
& $Depth (\mu m)$ & & & $Width (\mu m)$ & & & $\gamma$ & & & $\Lambda_1 (\mu m)$ & & & $\Lambda_2 (\mu m)$ & \\
& & & & & & & & & & & & & &\\
\hline & & & & & & & & & & & & & &\\
& $12$ & & & $12$ & & & 0.9864 & & & 4.5755 & & & 3.6457 &\\
& $10$ & & & $10$ & & & 0.9866 & & & 4.5726 & & & 3.6440 &\\
& $8$ & &  & $8$ & & & 0.9876 & & & 4.5691 & & & 3.6421 &\\
& $6.5$ & & & $6.5$ & & & 0.9975 & & & 4.5672 & & & 3.6412 &\\
& & & & & & & & & & & & & &\\
\hline
\end{tabular}
\caption{This table represents the data acquired for variation of the degree of entanglement and the two poling periods accounting different values of depths and widths of the waveguide to generate the entangled state shown in Eq. \eqref{MP1_Eq20}.}
\label{MP1_Table3}
\end{table}
As inferred from Table \ref{MP1_Table3}, such an entangled state can be maximally entangled for appropriate choice of depth and widths of the waveguide. Fig. \ref{MP1_Fig4} represents the calculated spectra of the modes corresponding to signal wavelengths $\lambda_{s1o}$, $\lambda_{s1e}$, where $o$, $e$ are ordinary and extraordinary polarizations respectively. Similarly, Fig. \ref{MP1_Fig5} represents the calculated spectra of the SPDC process for the generation of entangled state represented in Eq. \eqref{MP1_Eq20} corresponding to idler wavelength variation.

It will be interesting to know how the value of $\gamma$ varies for a type-I down-conversion process. We leave this in context of future work, as to the best of our knowledge, both these entangled states will have the same purpose i.e. choosing a single-mode frequency as per requirement through only one measurement, which we have already shown in context of Table \ref{MP1_Table3} and Fig. \ref{MP1_Fig3}, \ref{MP1_Fig4}. Frequency entanglement has attracted a lot of attention in the past years, both in theory and experiments, which is why direct generation of such entangled states with widely different frequencies will turn out to be extremely useful to today's experiments involving energy-time entanglement.
\section{Conclusion\label{MP1_Sec4}}

In this paper, we have proposed a dual periodically poled substrate to generate a two-pair frequency entangled state using type-0 SPDC process in a PPLN waveguide. We also address the possibility of the generated state to be maximally entangled by performing numerical simulations, considering values for different parameters of the waveguide. We then take into account the polarization of the four modes, and using type-II SPDC process, perform simulations to find out that the output entangled state is more effective and interesting. Such schemes should be of prime interest to quantum optical communication technologies and experiments involving integrated optical circuits. 

\section{Acknowledgments}
\label{MP1_Sec5}
Aakash Warke would like to thank Ms. Akshata Rao, MSRIT, Bangalore, for her useful expertise of SkecthUp software that helped build the geometry of the considered waveguide.

\end{document}